\documentclass[prl,twocolumn,floatfix,groupedaddress]{revtex4}
\usepackage{graphicx}
\usepackage{amsmath}

\newcommand{\be}{\begin{equation}}
\newcommand{\ee}{\end{equation}}
\newcommand{\ben}{\begin{eqnarray}}
\newcommand{\een}{\end{eqnarray}}

\newcommand{\nd}{\noindent}
\begin{document}

\title{Variational Principle underlying Scale Invariant Social Systems}

\author{A. Hernando$^1$, A. Plastino$^{2,\,3}$}
\address{$^1$ Laboratoire Collisions, Agr\'egats,
R\'eactivit\'e, IRSAMC, Universit\'e Paul Sabatier
118 Route de Narbonne 31062 - Toulouse CEDEX 09, France\\
$^2$National University La Plata, IFLP-CCT-CONICET, C.C. 727, 1900
La Plata, Argentina   \\ $^{3}$ Universitat de les Illes Balears
and IFISC-CSIC, 07122 Palma de Mallorca, Spain}

\begin{abstract}
MaxEnt's variational principle, in conjunction with
Shannon's logarithmic information measure, yields only exponential
functional forms in straightforward fashion.  In this
communication we show how to overcome this limitation via the
incorporation, into the variational process, of suitable dynamical
information. As a consequence, we are able to formulate a somewhat
generalized Shannonian Maximum Entropy approach  which provides a
unifying ``thermodynamic-like" explanation for the scale-invariant
phenomena observed in social contexts, as city-population
distributions. We confirm the MaxEnt predictions by means of
numerical experiments with random walkers, and compare them with
some empirical data.
\end{abstract}
\maketitle

% 89.70.Cf  ------- Entropy and other measures of information
% 05.90.+m  ------- Other topics in statistical physics, thermodynamics, and nonlinear dynamical systems
% 05.45.Df  ------- Fractals
% 02.50.-r  ------- Probability theory, stochastic processes, and statistics
%89.75.-k Complex systems
%89.75.Da Systems obeying scaling laws
%89.75.Fb Structures and organization in complex systems
%89.75.Hc Networks and genealogical trees
%89.75.Kd Patterns

\date{\today}

\section{Introduction}

Scale-invariant phenomena are plentiful in Nature and in
artificial systems.
%%CC%%
Illustrations can be encountered that range from
physical and biological to technological and social
sciences~\cite{uno}. Among the latter, one may speak of
empirical data from
%%CC%%
scientific collaboration networks~\cite{cites}, cites of physics
journals~\cite{nosotrosZ}, the Internet traffic~\cite{net1}, Linux
packages links~\cite{linux}, popularity of chess
openings~\cite{chess}, as well as  electoral
results~\cite{elec1,ccg}, urban
agglomerations~\cite{ciudad,ciudad2} and firm sizes all over the
world~\cite{firms}. These systems lack  a typical size, length or
frequency for observables under scrutiny, a fact that usually
leads to power law distributions
\begin{equation}\label{eq0}
p_X(x)\sim1/x^{1+\lambda},
\end{equation}
with $\lambda\geq0$. The class of universality defined by
$\lambda=1$,  corresponding to the so-called  Zipf's law (ZL) in
the cumulative distribution or the rank-size distribution received
much attention
~\cite{zipf,nosotrosZ,net1,linux,chess,ciudad,ciudad2,firms,chin,upf,citis}.
Maillart et al.~\cite{linux} have found that  links' distributions
follow ZL as a consequence of stochastic proportional growth. Such
kind of growth assumes that an element of the system becomes
enlarged proportionally to its size $k$, being governed by a
Wiener process. The $\lambda=1-$class emerges from a condition of
stationarity  (dynamic equilibrium)~\cite{citis}. ZL also applies
for processes involving either self-similarity~\cite{chess} or
fractal hierarchy~\cite{chin}, all of them mere examples amongst
very general stochastic ones~\cite{upf}. The instance $\lambda=0$
is representative of a second universality-class following Costa
Filho et al.~\cite{elec1}, who studied vote-distributions in
Brazil's electoral results. Therefrom emerge multiplicative
processes in complex networks~\cite{ccg}. $\lambda=0-$behavior
 ensues as well in i) city-population rank
distributions~\cite{nosotros}, ii) Spanish electoral
results~\cite{nosotros}, and iii) the degree distribution of
social networks~\cite{nosotros2}. As shown in Ref.
\cite{benford2}, this universality class encompasses Benford's
Law~\cite{benford1}. In the present vein, still another kind of idiosyncratic
distribution is often reported:  the log-normal one~\cite{lnwiki},
that has been observed in biology (length and sizes of living
tissue~\cite{bio}), finance (in particular the Black – Scholes
model~\cite{black}), and firms-sizes. The latter instance obeys
Gibrat's rule of proportionate growth~\cite{gibrat}, that also
applies to cities' sizes. \emph{Remarkably enough, all  these
variegated and sometimes quite complex systems share a scale-free
growth behavior}.

\nd Together with geometric Brownian motion, there is a variety of
models arising in different fields that yield Zipf's law and other
power laws on a case-by-case
basis~\cite{ciudad,ciudad2,citis,mod1,exp,renorm}, as preferential
attachment~\cite{net1} and competitive cluster growth~\cite{ccg}
in complex networks, used to explain many of the scale-free
properties of social, technological and biological networks. For
instance we may mention urban dynamics~\cite{ud}, opinion
dynamics~\cite{oppi}, and electoral results~\cite{elec1,elec2},
that develop detailed realistic approaches.  Ref.~\cite{otros} is
highly recommendable as a primer on urban modelling. Of course,
the renormalization group is intimately related to scale
invariance and associated techniques have been fruitfully
exploited in these matters (as a small sample see
\cite{renorm,voteGalam}).

\nd Can the celebrated maximum entropy principle of Jaynes'
(MaxEnt) \cite{jaynes,katz} explain the disparate phenomena
recounted above? At first sight, one faces a seemingly
unsurmountable difficulty. MaxEnt's variational principle, in
conjunction with Shannon's logarithmic information measure, can
not {\it in straightforward fashion} yield power laws (nor more
involved combinations that include power laws), but only
exponential distributions~\cite{katz}. This fact constituted one
major motivation~\cite{pla} for introducing Tsallis' information
measure~\cite{tbook} (and its associated Tsallis-MaxEnt
treatment~\cite{pp93}). One immediately gets power laws thereby as
a result of varying the measure.
%%%
We have shown in Ref.~\cite{ourRef} how to overcome the above
Shannonian limitation via suitable incorporation, into the MaxEnt
principle, of  dynamical information.
%%%
As a consequence, we are now able to formulate a Shannonian
Maximum Entropy approach~\cite{jaynes,katz} common to {\it all}
these systems. As stated in~\cite{ourRef}, this technique exhibits
the peculiarity of including ``equations of motion" as a part of
the required a priori knowledge which always needs to be
incorporated into the accompanying Jaynes' variational problem.
The desired goal can in this way  be successfully achieved, which
provides a unifying ``thermodynamic-like" explanation for the
above mentioned disparate phenomena. The approach explicitly
reconciles two apparent different viewpoints, those attached to i)
growth models and ii) information-treatments \cite{X2}.

\section{Theoretical procedure}

\subsection{General approach}

Accordingly, consider a rather general stochastic dynamical equation for
the observable $x$, namely,
\begin{equation}\label{eq1}
\dot{x}(t) = k \,g[x(t)];\,\,g\,\,{\rm arbitrary}.
\end{equation}
where $k$ represents the derivative of a Wiener process
(characterized by stationary independent increments) that
frequently occurs in economic and social systems. We assume that
there exists an appropriate transformation of $x$ that makes
(\ref{eq1}) invariant and a hallmark of the \emph{symmetry}
characterizing the system. Thermodynamics and many areas of
physics have been shown by Frieden and Soffer \cite{FS} to be
typified by \emph{translational} symmetry if they are
theoretically described \`a la Fisher \cite{FS}. This entails
$g(x)=1$, which makes Eq.~(\ref{eq1}) the emblematic equation for
linear Brownian motion. Our considerations will revolve around a
new variable $u=u(x)$ such that $dx/du=g(x)$, a variable that
\begin{enumerate} 
\item should linearize the dynamic equation, 
\item make the original $x-$symmetry a translational one $\dot{u}(t) = k$, and
\item according to the ``central hypothesis" of Ref. \cite{ourRef},
constitutes the tool for introducing dynamical information into
MaxEnt by working with the $u-$entropy of the system.
\end{enumerate} One has \cite{ourRef}
\begin{equation}
S[p_U(u)] = \int_\Omega du p_U(u)\log[p_U(u)],
\end{equation}
with $\Omega$ an appropriate ``volume'' in $u-$space. The equilibrium probability
density $p_U(u)$ is derived from Jaynes' MaxEnt principle\cite{jaynes}
\begin{equation}
\delta_{p}\left\{S[p_U(u)] - \sum_i \lambda_i \langle f_i(u)\rangle\right\}=0,
\end{equation}
where the mean values of the functions $f_i(u)$ describe the
a-priori  known  constraints on the system, while $\lambda_i$ are
the associated Lagrange multipliers.  Our constraints represent
{\it conservation rules}, operating on our system, that strongly
condition the configuration of the equilibrium state. The general
solution of the Jaynes problem  is \cite{jaynes}
\begin{equation}\label{equ}
p_U(u)du = \exp{\left[-\sum_i \lambda_i f_i(u)\right]}du
\end{equation}
which, changing back to  $x$ lead us to a $p_X(x)$ of the form
\begin{equation}\label{eqx}
p_X(x)dx = \exp{\left[-\sum_i \lambda_i f_i(u(x))\right]}\frac{dx}{g(x)}.
\end{equation}
The main {\it difference}  with the {\it usual} MaxEnt (ME)
solution is the Jacobian $du/dx=1/g(x)$. This looks trivial
enough, but {\it the Jacobian contains dynamical information},
otherwise absent from the treatment. In a manner of speaking, {
we are thereby ``extending" the exponential-like form of the
Jaynes' ME-solutions to other analytical possibilities}.

\subsection{Scale-invariant systems}

To illustrate this procedure we consider the emblematic equation used
in many models of mathematical finance (e.g., the Black-Scholes
model~\cite{black}) and cities' and firms' sizes (Gibrat's
law~\cite{lognorm}), and also many of the ``social" examples
listed above. We speak then of the so-called geometrical Brownian
motion:
\begin{equation}\label{eqPG}
\dot{x}(t) = k x(t),
\end{equation}
which is symmetric under scale transformations $x'=cx$ with $c$ an
arbitrary constant. We are here defining a proportional growth or
multiplicative process. A systems following such  dynamics can be
described by our approach. We set $u=\log(x/x_0)$, where $x_0$ is
the minimum allowed value for $x$ (or a ``reference"-one). The
Jacobian of the transformation is $du/dx=1/x$, and the volume in
$u$ space is defined as $\Omega$: $[0\leq u\leq u_M]$. Here
$u_M=\log(x_M/x_0)$ corresponds to some maximum allowable value
(which can be infinity).
%%CC%%>
When no constraints are included  and an infinite volume $\Omega$
is considered, the constituents of system diffuse as random
walkers in $u$. Accordingly, the density distribution (DD)
$p_U(u)du$ is a gaussian distribution that becomes a log-normal
one in $x$. This no-constraint case has been widely studied, and most
city-population distributions follow it \cite{lognorm}. 

Let us discuss now just how adding {\it constraints} affect the
equilibrium DD by
%%CC%%<
 considering the simple form $f_i(u) = u^i$.
%%CC%%
The case $i=0$ corresponds to normalization of the DD, while $i=1$
 refers  to  having  $\langle u\rangle$ as a constraint. We tackle the
variational solution subjected to these two constraints, setting
$\mu=\lambda_0$ and $\lambda=\lambda_1$. Accordingly, the
equilibrium DD $p_U(u)$  extremizes the functional $F = S - \mu -
\lambda\langle u\rangle,$ so that
\begin{equation}
p_U(u)du = e^{-\mu-\lambda u}du.
\end{equation}
The values for $\mu$ and $\lambda$ are obtained from the
conservation rules $1 = \int_\Omega du~e^{-\mu-\lambda u}$ and
$\langle u\rangle = \int_\Omega du~e^{-\mu-\lambda u}u$
which yield $\langle u\rangle=e^\mu=1/\lambda$ if
$u_M\rightarrow\infty$. The DD, as a function of $x$, becomes
\begin{equation}
p_X(x)dx  = e^{-\mu}x_0^\lambda\frac{dx}{x^{\lambda+1}},
\end{equation}
i.e., a power law
%%CC%%
whose exponent is characterized by $\lambda$, as that presented in
Eq. (\ref{eq0}). Look first at the particular solution $\lambda=0$
(no constraint on $\langle u\rangle$). One is led to $e^{\mu}=u_M$
and $\langle u\rangle=u_M/2$. In terms of $x$ the DD is
\begin{equation}\label{eqBL}
p_X(x)dx = \frac{1}{u_M}\frac{dx}{x}.
\end{equation}
As remarked above, such  DD is related to Benford's Law. Using a
thermodynamic analogy, the law describes the simple scenario of a
non-interacting system confined to  a finite volume of $u$-space,
with a Gaussian distribution for $\dot{u}$, i.e., a scale-free
ideal gas (SFIG)\cite{nosotros}.
%%CC%%
Even if the normalization seems to diverge here for
$u_M\rightarrow \infty$, it can be kept finite in going to the
thermodynamic limit (see \cite{nosotros}): if $N$ is the total
number of system's ``elements", its density $\rho(x)=Np_X(x)$ is
normalized in the limit ($N,\,u_M\rightarrow \infty$) if $N/u_M
\rightarrow \rho_0$, where $\rho_0$ is a constant. Let us pass now
to the particular solution $\mu=0$ (relaxing the normalization
constraint). If $u_M\rightarrow \infty$, we obtain from the
conservation rules $\lambda=1$ and $\langle u\rangle=1$.
Accordingly, the DD for the observable $x$ is
\begin{equation}\label{eqZ}
p_X(x)dx  = x_0\frac{dx}{x^2},
\end{equation}
i.e., Zipf's law (ZL). In a thermal context, the absence of
normalization can be understood as an inability of the system to
reach the thermodynamic limit. One may then speak of a  Zipf
regime \cite{nosotros}. The lack of  normalization constraint is
discussed also in  Refs.~\cite{nosotrosZ,upf}. An
 explanation may be concocted:  these
elements might be distributed on the surface of an appropriate
volume \cite{nosotrosZ,nosotros}. Accordingly, ZL usually holds
for the upper tail and not for  the ``bulk'' of the associated
distribution.
%%CC%%
There is a free interchange of elements between the two regions
with no ``energetic cost'' since the ``chemical potential'' $\mu$
is zero. Accordingly, the number of elements does not remain
constant (photon statistics). This number-fluctuation  has no
effect in the bulk region where the thermodynamic limit is
reached, but it totally determines how the density behaves on our
putative ``surface".
%%CC%%

\section{Present results: numerical experiments and empirical observations}

So as to confirm the preceding MaxEnt predictions we have
performed numerical experiments with random walkers and  compared
them with empirical city-population data. We solve numerically the
equation $\dot{x}(t)=kx(t)$, discretizing the time in intervals of
$\Delta t$.  $k$ is randomly generated at each iteration from a
Gaussian distribution with zero mean and variance $K$.

%%%%%%%%%%%%%%%%%%%%%%%%%%%%%%%%%%%%%%%%%
\begin{figure}[t]
\begin{center}
\includegraphics[width=0.8\linewidth]{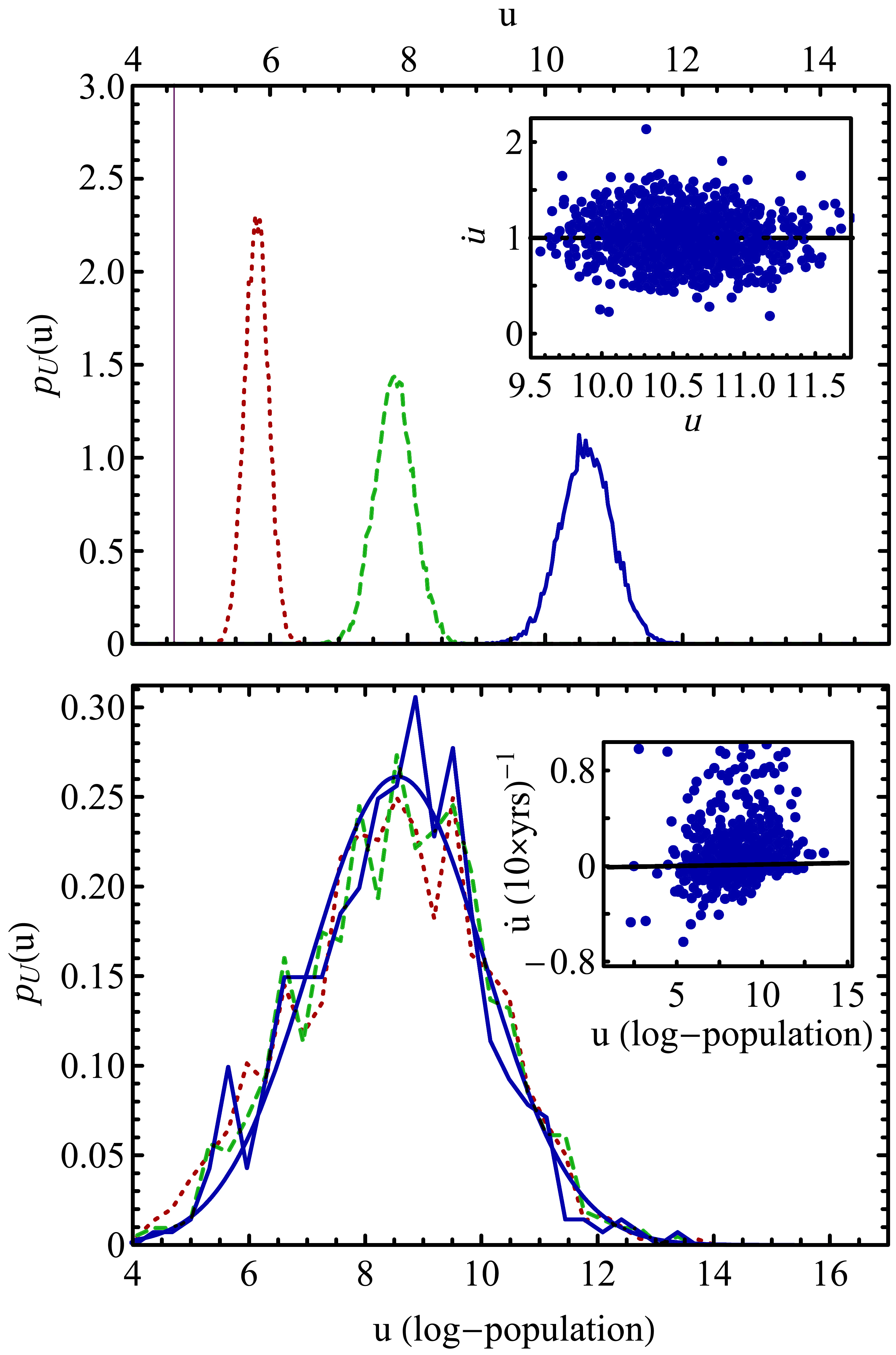}
\caption[]{Top panel: Evolution of a random walkers' distribution
without constraints from an initial delta-one (purple solid line),
passing through intermediate stages (red dotted, green dashed, and
solid blue lines). Inset: logarithmic size $u$ vs. relative growth
$\dot{u}$ (linear fit in black). Bottom panel: Florida's
cities-population-distribution for 1990, 2000, and 2010 (same
color-code as TP) compared with a normal distribution (smooth
solid blue line). Inset: same as in top panel's inset.}
\end{center}
\end{figure}
%%%%%%%%%%%%%%%%%%%%%%%%%%%%%%%%%%%%%%%%%

\subsection{Free evolution}

We first study  evolution without constraints (as a control case),
starting with  $N=10000$ walkers located at the same position
$x=100$ ($u=2\log10$), that evolve with $K=10$ and a time interval
$\Delta t=10^{-5}$ in $x$. We also include a drift, as in
Ref.~\cite{gibrat}. As expected, we obtain the diffusion process
depicted in upper panel of Fig. 1, which is a log-normal DD in $x$
(and a Gaussian in $u$).

\nd We now compare such evolution with that of the
cities-population of Florida State using
 Census Bureau's data for the years 1990, 2000,  and 2010 \cite{usa}, also
finding a log-normal in $x$ with growing width (1.64, 1.65 and
1.72, respectively). The proportional growth condition is checked
by computing the correlation between the logarithm of the
population of each place $u$ with its relative change $\dot{u}$,
using two-points formulas for the time derivative. For $\dot{u}=k$
both observables are independent for scale-free systems. The
correlation value is 0.027, small enough to confirm that
geometrical motion is taking place (see inset of Fig. 1, bottom
panel). Even if the populations of Florida-cities do not seem to
match the non-interacting assumption of our simulation, we expect
a short-range nature for the correlation between the population of
each pair of cities   \cite{Gmodel}, relative to the State's size.
According to our thermodynamic interpretation, we can think of a
dilute scale-free gas at zero pressure and expanding freely.

\subsection{Evolution constrained by normalization and finite volume}

For an example with normalization constraint and finite volume
($\lambda=0$), we arbitrarily define $x_0=1$, $x_M=10^4$,
($u_M=4\log10$) and use the same initial conditions as in the
precedent case. Now, a ``move" is not accepted if the new position
falls outside the appropriate region. After some iterations the
system reaches equilibrium, as shown in Fig. 2, top panel. We find
a nice fitting of this equilibrium distribution to that predicted
by MaxEnt [Eq.~(\ref{eqBL})], confirming the validity of our
approach.

\nd A  system in this thermodynamical condition
\begin{enumerate} \item obeys the proportional growth dynamics, \item exhibits low
correlation between its elements, \item conserves the
particle-number, and \item is characterized by  an objective,
measurable size-cons\-traint. \end{enumerate}This last condition, for
city-populations,  has the form of a geographical constraint. Such
is the case of, e.g., the Marshall Islands: this particular
geographical area covers 181.3 km$^2$ distributed into 29 atolls
and 5 islands. Traversing the sea may reduce correlations between
cities as compared to {\it terra firma}. The migration pattern
concentrates in the two main population-centers, Rita and Ebeye,
so that only indirect correlations are expected between the rest
of the cities. We verified this issue by recourse to data from
1980, 1988, and 1999 \cite{mar}. The islands' logarithmic
population $u$, exhibits a correlation coefficient of
$9\times10^{-5}$ with a relative increment $\dot{u}$, confirming
the dynamics' nature (inset of Fig. 1b, bottom panel). We  fitted
the raw data for  all low-correlated centers (154 populations) to
the MaxEnt prediction and also to a log-normal with the same
log-mean and log-variance that characterize the data (Fig. 1b,
bottom panel). We found a correlation of $0.991$ in the former
case and of $0.979$ in the latter one. One thus may visualize the
Marshall Islands as a closed-volume scale-free ideal ga (SFIG) in
thermal equilibrium, again empirically confirming the validity of
our approach.

%%%%%%%%%%%%%%%%%%%%%%%%%%%%%%%%%%%%%%%%%
\begin{figure}[t]
\begin{center}
\includegraphics[width=0.82\linewidth]{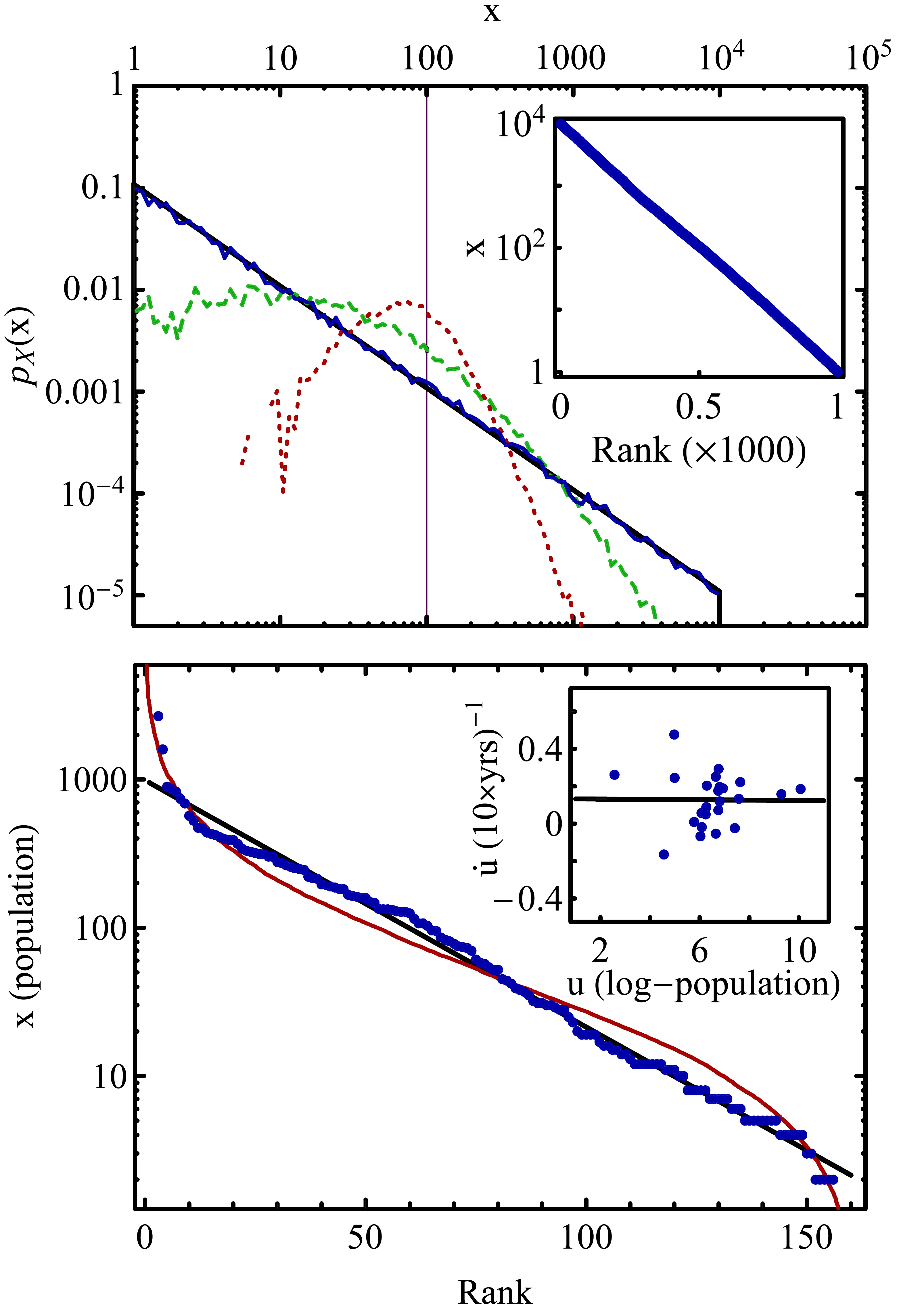}
\caption[]{ Top panel: walkers' distribution for the first
text-example $\lambda=0$ (origin of Benford Law). The blue solid
line indicates convergence towards the MaxEnt-prediction (black
solid line). Inset: equilibrium rank-distribution. Bottom Panel:
rank-distribution of Marshall Islands' city-population versus i)
MaxEnt-prediction for $\lambda=0$ (black line) and ii) a
log-normal (red line). Inset: logarithmic size $u$ vs. relative
growth $\dot{u}$ (linear fit in black). }
\end{center}
\end{figure}
%%%%%%%%%%%%%%%%%%%%%%%%%%%%%%%%%%%%%%%%%

%%%%%%%%%%%%%%%%%%%%%%%%%%%%%%%%%%%%%%%%ACA ACA ACA

\subsection{Evolution constrained by a mean value-condition}

\nd   The second example presented here ($\mu=0$) uses a simple
algorithm that reproduces Zipf's Law. $N$ walkers ``unfold" while
guaranteeing the conservation laws' proper working: $\langle
\log(x/x_0)\rangle=\langle u\rangle=1$ (or
$\sum_{i=1}^N\log(x_i/x_0)=\sum_{i=1}^N u_i=N$). We start all the
walkers at $x=e\times x_0$ (any distribution with $\langle
u\rangle=1$ is adequate) and proceed iterating as follows:
\begin{enumerate} \item  select an arbitrary walker and change
its position in proportional fashion according to $x(t+\Delta
t)=x(t)\times(1+k\Delta t)$, with a random value $k$;  \item 
assume that we have a bulk reservoir so that the exchange of
elements has no 'energetic cost'. We randomly select a second
walker and remove him from his position $x'$; \item  add a
new walker at a position that preserves the operating conservation
rule, that is $x'(t+\Delta t)=x'(t)\times(1-k\Delta t)$.
\end{enumerate}
%%CC%%
We iterate these steps till convergence is achieved. After some
iterations, we do converge  to the predicted MaxEnt distribution
Eq.~(\ref{eqZ}) (Fig. 3, top panel, with the same values for
$x_0$, $N$, $K$, and $\Delta t$ used in the first example). thusly
Once more, we reconfirm that the approach presented here works in
reasonable fashion. Remarkably enough, a similar algorithm is also
able to reproduce {\it any arbitrary power law with exponent}
$\lambda+1$ by changing the value of the conservation rule in the
fashion $\langle u\rangle=1/\lambda$.

\nd         For city-populations,  elements-exchange  occurs in a
continuous fashion. However the data are recorded at intervals of
years, so that the ensuing exchange-effects are similar to those
of our simulation. Areas belonging to the most-populated sites
eventually change with time. Thus, we deal with an open system
with no fixed number of elements, like our photon-gas above. As a
well known example consider the most populated metropolitan
USA-areas \cite{citis}. We have confirmed the proportional growth
hypothesis using data from years 1990, 2000, and 2010 \cite{usa}
to find a small correlation of 0.016 between $u$ and $\dot{u}$
(inset of Fig. 1c, bottom panel). Our first $N=150$ areas are in
Zipf's regime (Fig. 3, bottom panel) and constitute the surface of
the statistical system at hand. We have verified the prevailing
conservation rule, finding
$\sum_{i=1}^{150}\log(x_i/x_{150})=145.7$, 150.8, and 154.9 for
each year, respectively, close to the MaxEnt prediction of 150. We
have also verified that 10 of these areas (a 6.7\%) that in 1990
pertained to the Zipf regime are not characterized by it in 2010.

%%%%%%%%%%%%%%%%%%%%%%%%%%%%%%%%%%%%%%%%%
\begin{figure}[t]
\begin{center}
\includegraphics[width=0.8\linewidth]{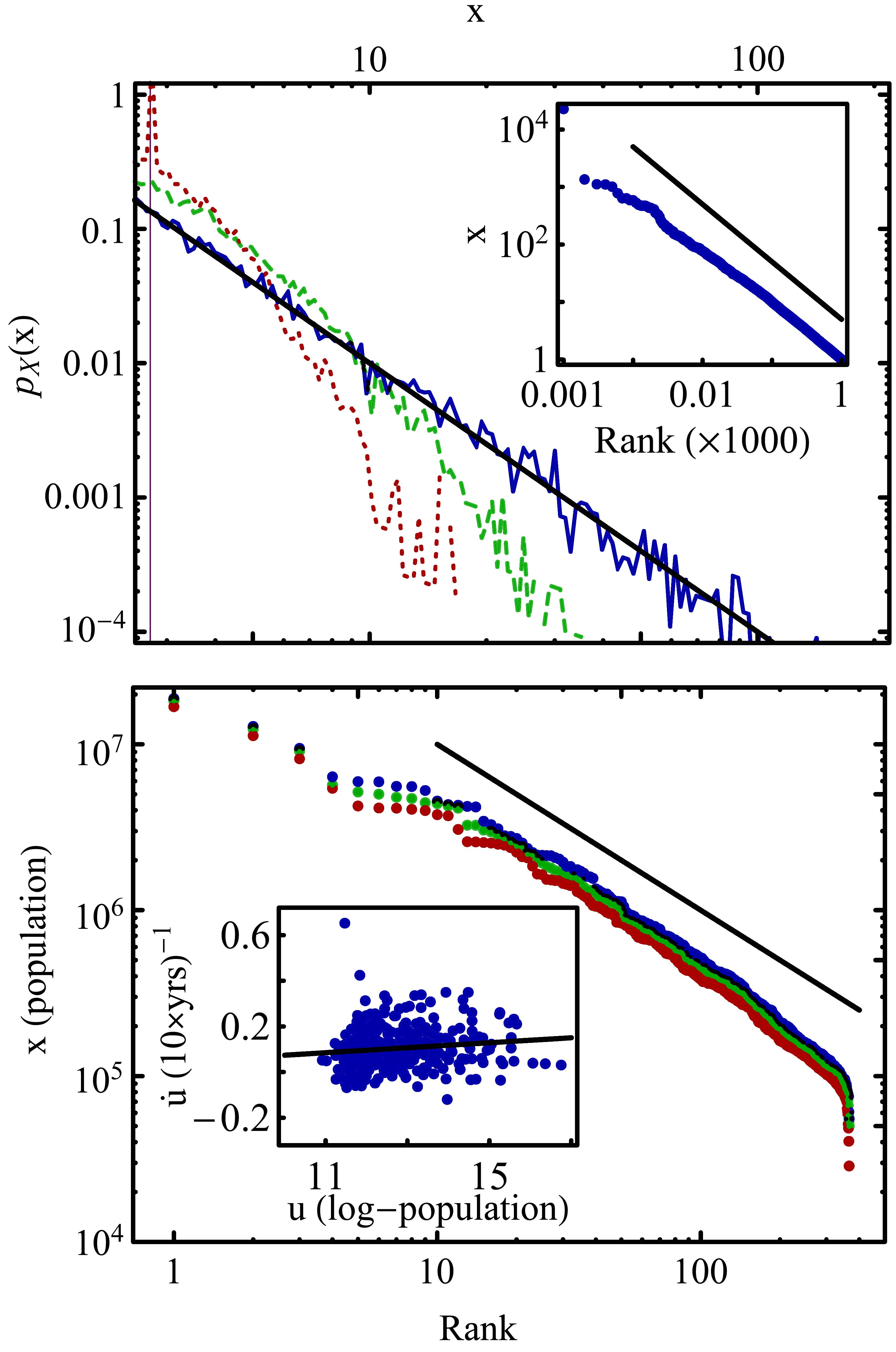}
\caption[]{Top panel: walkers distribution
for the second text-example $\mu=0$ (Zipf's Law). Inset: equilibrium
rank-distribution (slope$=1$ in black). Bottom panel: rank-distribution of USA's
most-populated metropolitan areas for 1990, 2000, and 2010; Inset: logarithmic
size $u$ vs. relative growth $\dot{u}$ (linear fit in black).}
\end{center}
\end{figure}
%%%%%%%%%%%%%%%%%%%%%%%%%%%%%%%%%%%%%%%%%

\section{Conclusions}

Different phenomena involving scale-invariance can be unified via
Jaynes' MaxEnt principle, provided that adequate dynamical
information is suitably incorporated into the variational process,
in the manner here prescribed. This allows one to perform a
thermodynamic-like description of social systems, related to that of an ideal gas.\\

\vfill
{\bf ACKNOWLEDGMENT:}  This work was partially supported by
ANR DYNHELIUM (ANR-08-BLAN-0146-01) Toulouse, project PIP1177 of CONICET,
(Argentina), project FIS2008-00781/FIS (MICINN), and project FEDER (EU) (Spain, EU).

\end{document}